\documentclass[final]{svjour2}
\usepackage{graphicx}
\usepackage{rotating}
\usepackage{amssymb}
\usepackage{mathptmx}
\usepackage[numbers,sort&compress]{natbib}
\usepackage{lipsum}
\usepackage{subfig}
\usepackage{sidecap}
\usepackage{wasysym}
\usepackage[font={small}]{caption}
\usepackage{bm}
\usepackage{tabularx}
\usepackage[hang,flushmargin,bottom]{footmisc} 
\makeatletter
\newcommand\subparagraph{%
  \@startsection{subparagraph}{5}
  {\parindent}
  {3.25ex \@plus 1ex \@minus .2ex}
  {-1em}
  {\normalfont\normalsize\bfseries}}
\makeatother
\makeatletter
\def\@maketitle{\newpage
\normalfont
\vbox to0pt{\if@twocolumn\vskip-39pt\else\vskip-49pt\fi
\nointerlineskip
\makeheadbox\vss}\nointerlineskip
\vbox to 0pt{\offinterlineskip\rubricwidth=\columnwidth
\vskip-12.5pt
\if@twocolumn\else 
   \divide\rubricwidth by144\multiply\rubricwidth by89 
   \vskip-\topskip
\fi
\hrule\@height0.35mm\noindent
\advance\fboxsep by.25mm
\global\advance\rubricwidth by0pt
\rubric
\vss}\vskip-42pt 
\if@twocolumn\else
 \gdef\footnoterule{%
  \kern-3\p@
  \hrule\@width\columnwidth  
  \kern2.6\p@}
\fi
 \setbox\authrun=\vbox\bgroup
     \hrule\@height 9mm\@width0\p@
     \pretolerance=10000
     \rightskip=0pt plus 4cm
    \nothanksmarks
    {\authorfont
    \setbox0=\vbox{\setcounter{auth}{1}\def\and{\stepcounter{auth} }%
                   \hfuzz=2\textwidth\def\thanks##1{}\@author}%
    \setcounter{footnote}{0}%
    \global\value{inst}=\value{auth}%
    \setcounter{auth}{1}%
    \if@twocolumn
       \rightskip43mm plus 4cm minus 3mm
    \else 
    \fi
\def\and{\unskip\nobreak\enskip{\boldmath$\cdot$}\enskip\ignorespaces}%
    \noindent\ignorespaces\@author\vskip3.62pt} 
    {\LARGE\bfseries
     \noindent\ignorespaces
     \@title \par}\vskip 1.62pt\relax 
    \if!\@subtitle!\else
      {\large\bfseries
      \pretolerance=10000
      \rightskip=0pt plus 3cm
      \vskip-5pt
      \noindent\ignorespaces\@subtitle \par}\vskip 0.24pt
    \fi
    \small
    \if!\@dedic!\else
       \par
       \normalsize\it
       \addvspace\baselineskip
       \noindent\@dedic
    \fi
 \egroup 
 \@tempdima=\headerboxheight
 \advance\@tempdima by-\ht\authrun
 \unvbox\authrun
 \ifdim\@tempdima>0pt
    \vrule width0pt height\@tempdima\par
 \fi
 \noindent{\small\@date\vskip -2.24pt} 
 \global\@minipagetrue
 \global\everypar{\global\@minipagefalse\global\everypar{}}%
}
\makeatother

\usepackage{titlesec}
\let\subparagraph\relax 
\usepackage{titlesec}
\makeatletter
\journalname{Journal of Low Temperature Physics}
\titlespacing\section{0pt}{12pt plus 4pt minus 10pt}{0pt plus 2pt minus 10pt}
\titlespacing\subsection{0pt}{12pt plus 4pt minus 10pt}{0pt plus 2pt minus 10pt}
\titlespacing\subsubsection{0pt}{12pt plus 4pt minus 10pt}{0pt plus 2pt minus 10pt}
\bibpunct{}{}{,}{s}{}{,}
\newcommand*{\doi}[1]{DOI:#1}

\newcommand{\EE}{\emph{E\kern0.75ptE} }
\newcommand{\TT}{\emph{T\kern0.75ptT} }
\newcommand{\BB}{\emph{B\kern0.75ptB} }

\begin{document}
\setcitestyle{numbers,square}

\newcommand{\hdblarrow}{H\makebox[0.9ex][l]{$\downdownarrows$}-}
\title{Magnetic Sensitivity of AlMn TESes and Shielding Considerations for Next-Generation CMB Surveys}

\author{
E.M.~Vavagiakis\textsuperscript{1}\kern-1.5pt \and           
S.W.~Henderson\textsuperscript{1}\kern-1.5pt \and            
K.~Zheng\textsuperscript{1}\kern-1.5pt \and                  
H.-M.~Cho\textsuperscript{2}\kern-1.5pt \and
N.F.~Cothard\textsuperscript{1}\kern-1.5pt \and
B.~Dober\textsuperscript{3}\kern-1.5pt \and
S.M.~Duff\textsuperscript{4}\kern-1.5pt \and
P.A.~Gallardo\textsuperscript{1}\kern-1.5pt \and
G.~Hilton\textsuperscript{4}\kern-1.5pt \and
J.~Hubmayr\textsuperscript{4}\kern-1.5pt \and
K.D.~Irwin\textsuperscript{2,5}\kern-1.5pt \and
B.J.~Koopman\textsuperscript{1}\kern-1.5pt \and
D.~Li\textsuperscript{2}\kern-1.5pt \and
F.~Nati\textsuperscript{6}\kern-1.5pt \and
M.D.~Niemack\textsuperscript{1}\kern-1.5pt \and
C.D.~Reintsema\textsuperscript{3}\kern-1.5pt \and
S.~Simon\textsuperscript{7}\kern-1.5pt \and
J.R.~Stevens\textsuperscript{1}\kern-1.5pt \and
A.~Suzuki\textsuperscript{8}\kern-1.5pt \and
B.~Westbrook\textsuperscript{9}\kern-1.5pt
}

\institute{\footnotesize
  \noindent\textsuperscript{1}Department of Physics, Cornell University, Ithaca, NY, USA 14853\\
  \noindent\textsuperscript{2}SLAC National Accelerator Laboratory, 2575 Sandy Hill Road, Menlo Park, CA, USA 94025\\
  \noindent\textsuperscript{3}NIST Quantum Devices Group, 325 Broadway Mailcode 817.03, Boulder, CO, USA 80305\\
  \noindent\textsuperscript{4}National Institute of Standards and Technology, Boulder, CO, USA 80305\\
  \noindent\textsuperscript{5}Department of Physics, Stanford University, Stanford, CA, USA 94305-4085\\
  \noindent\textsuperscript{6}Department of Physics and Astronomy, University of Pennsylvania, Philadelphia, PA, USA 19104\\
  \noindent\textsuperscript{7}Department of Physics, University of Michigan, Ann Arbor, MI, USA 48103\\
  \noindent\textsuperscript{8}Lawrence Berkeley National Laboratory, Berkeley CA, USA 94720\\
  \noindent\textsuperscript{9}Department of Physics, University of California, Berkeley, CA, USA 94720\\  
  \email{ev66@cornell.edu}}

\maketitle

\begin{abstract}
In the next decade, new ground-based cosmic microwave background (CMB) experiments such as Simons Observatory, CCAT-prime, and CMB-S4 will increase the number of detectors observing the CMB by an order of magnitude or more, dramatically improving our understanding of cosmology and astrophysics. These projects will deploy receivers with as many as hundreds of thousands of transition edge sensor (TES) bolometers coupled to superconducting quantum interference device (SQUID)-based readout systems. It is well known that superconducting devices such as TESes and SQUIDs are sensitive to magnetic fields. However, the effects of magnetic fields on TESes are not easily predicted due to the complex behavior of the superconducting transition, which motivates direct measurements of the magnetic sensitivity of these devices. We present comparative four-lead measurements of the critical temperature versus applied magnetic field of AlMn TESes varying in geometry, doping, and leg length, including Advanced ACT and POLARBEAR-2/Simons Array bolometers. MoCu ACTPol TESes are also tested and are found to be more sensitive to magnetic fields than the AlMn devices. We present an observation of weak-link-like behavior in AlMn TESes at low critical currents. We also compare measurements of magnetic sensitivity for time division multiplexing SQUIDs and frequency division multiplexing microwave ($\mu$MUX) rf-SQUIDs. We discuss the implications of our measurements on the magnetic shielding required for future experiments that aim to map the CMB to near-fundamental limits. 

\end{abstract}

\section{\label{sec:intro}Introduction}
\vspace{1em}

Current cosmic microwave background (CMB) experiments depend on arrays of
superconducting transition edge sensor (TES) bolometers coupled to
superconducting quantum interference device (SQUID)-based readout systems to
make measurements of microwave wavelength photons. New generations of these technologies are being developed for upcoming experiments, including CCAT-prime, a 6-m
aperture off-axis submillimeter telescope that will be located at 5600 m
elevation on Cerro Chanjnantor in Chile \citep{ccatp}, and Simons Observatory \citep{simonsobs}, an array of
new CMB telescopes that will be located at 5200 m elevation on Cerro
Toco in Chile, near the Atacama Cosmology Telescope (ACT) \citep{Fowler2006}, CLASS \citep{CLASS},  
and Simons Array \citep{Arnold2012}. These technologies are also relevant for CMB-S4, the
next-generation ground-based CMB project \citep{CMBS4_Science} which will enable tests of inflation and provide constraints on dark energy and fundamental particle physics.  

The superconducting devices on which current and future CMB surveys depend are sensitive to magnetic fields, and the response of the devices to external magnetic fields needs to be understood. When a magnetic field is applied to the plane of the superconducting film of a TES, for example, the critical temperature of the device shifts, which can affect the performance of the device in multiple ways. Sources of magnetic fields that can interfere with TESes and SQUIDs include Earth's DC field as well as AC fields produced by nearby instrumentation and the telescope's motion through Earth's field. If the devices are not shielded sufficiently, exposure to magnetic fields could result in the presence of artifacts in the CMB temperature and polarization maps that could negatively impact science goals and are difficult to remove. Device performance in the presence of magnetic fields is difficult to compute analytically, rendering direct measurements necessary to understand the behavior of these devices. Information about SQUID and TES magnetic sensitivity will motivate magnetic shielding design considerations for future CMB experiments.

One technique of fabricating TES bolometers uses thin films of aluminum doped with manganese impurities to reduce the $T_\mathrm{c}$ of the film from $\sim$1 K to $\sim$100 mK. This approach has advantages in the simplicity of fabrication and results in reduced sensitivity to magnetic fields when compared to MoCu bilayer fabrication techniques \citep{deiker2004}. AlMn TESes can also be fabricated on single wafers with high uniformity, as are currently being used for Advanced ACTPol (AdvACT) and POLARBEAR-2/Simons Array \citep{henderson2015,Duff2016}. These features along with demonstrated performance in the field make AlMn TESes an attractive choice for next generation CMB experiments. Multiplexing readout of TESes is currently achieved with either time division multiplexing (TDM) using DC SQUIDs \cite{CMBS4_Technology} or frequency division multiplexing (FDM) using MHz LC resonators \citep{Lanting2005} or rf-SQUIDs ($\mu$MUX) \citep{IrwinandLehnert,Mates2008,CMBS4_Technology}. In this work, AlMn TESes and MoCu bilayer TESes (from ACTPol), TDM DC SQUIDs, and FDM $\mu$MUX rf-SQUIDs are tested for magnetic sensitivity.

The treatment of magnetic shielding currently varies for CMB experiments. Instruments for ACT rely on TDM readout and have used multiple layers of Cryoperm and Amumetal 4K in combination with individual niobium shields for SQUID series arrays \citep{Thornton2008,Ward2016}. Experiments using MHz FDM readout systems, like POLARBEAR and SPT-3G, have mounted SQUIDs on Nb foil surrounded by a small cryoperm sleeve \citep{QuealyThesis}. Appropriate shielding factors will be motivated by experimental testing of these SQUIDs and TESes combined with simulated telescope observations. This information will be combined with simulations of shield geometries in order to develop mechanical designs for the cryogenic receivers currently under development. 

\section{\label{sec:section1}Magnetic Sensitivity of TES Critical Temperatures}
\vspace{1em}

We take resistance measurements of TESes varying in geometry, material, doping, manufacturer, and leg length using four-lead measurements, which precisely read out the low resistance values and critical temperatures by eliminating the lead and contact resistances from the measurements. TES chips were wire bonded and affixed with rubber cement to a printed circuit board stripped of solder and mounted to the coldest (100 mK) stage of a dilution refrigerator (DR). A set of 1-m-diameter Helmholtz coils applied DC magnetic fields up to 10.5 Gauss to the outside of the DR, with the fields applied perpendicular to the plane of the devices being tested. The fields were attenuated by a 30-cm-diameter, 85-cm-long half-open cylindrical room temperature mu-metal magnetic shield inside the DR. Shielding factors were measured by using a gaussmeter to measure the field between the coils with and without the shield in place and were determined to be 380 $\pm$ 20 with the axis of the coils perpendicular to the axis of the shield (for the SQUID and weak-link-like behavior measurements) and 2.9 $\pm$ 0.2 with the axis of the coils coincident with the axis of the shield (for the four-lead measurements) at the locations of our detectors. The series array modules used for TDM readout are additionally shielded in a niobium box. Resistance versus temperature data were acquired for each TES at various values of applied magnetic field, using a lakeshore AC resistance bridge with a low-noise preamplifier and ruthenium oxide thermometry with low magnetic field-induced errors (Fig. 1).  

\begin{figure}
  \centering
  \subfloat[]{\includegraphics[width=0.45\linewidth,keepaspectratio]{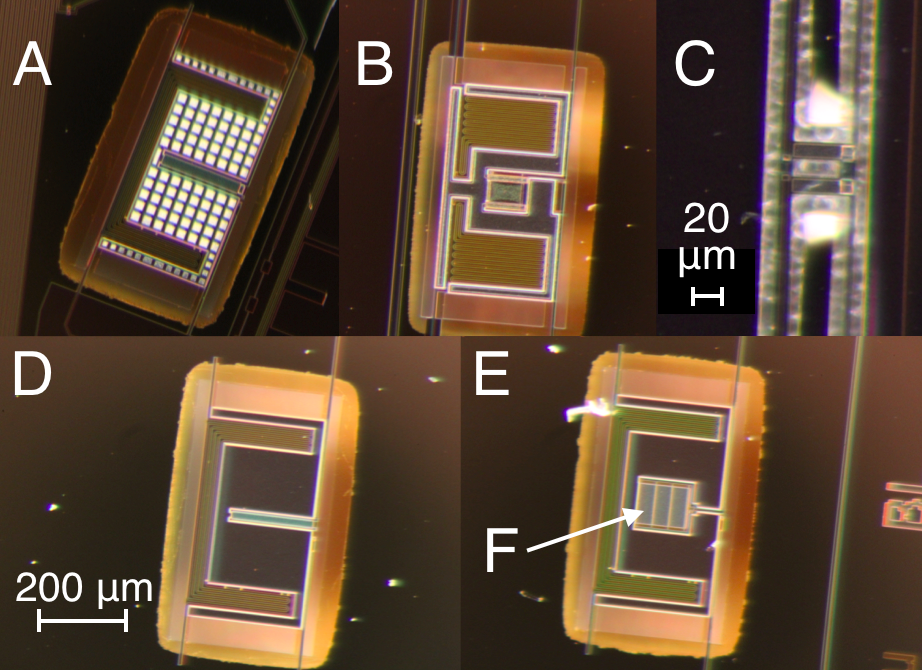}\label{fig:f1}}
  \hfill
  \subfloat[]{\includegraphics[width=0.54\textwidth]{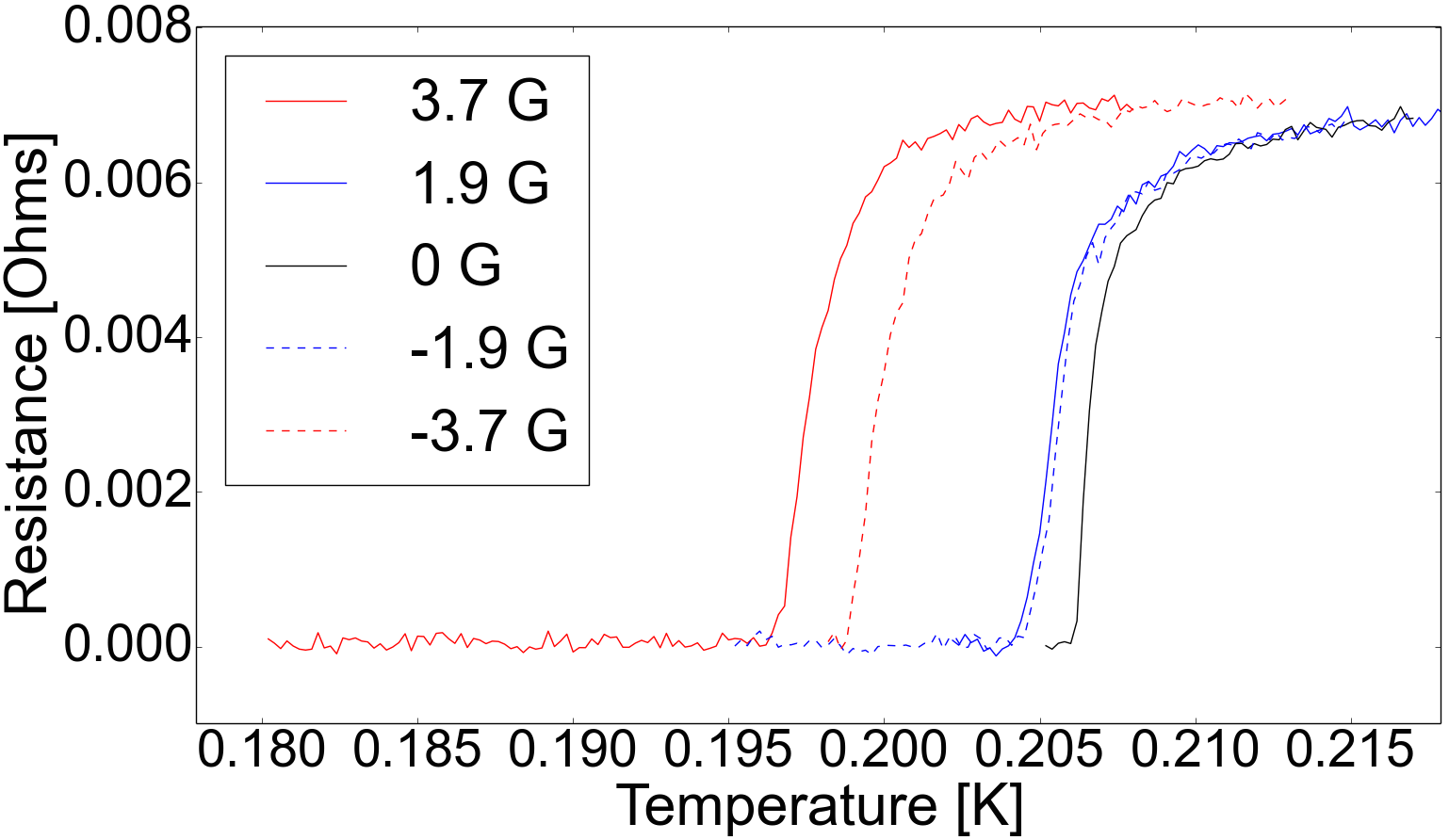}\label{fig:f1}}
  \caption{(a): Images of tested TESes. A: AdvACT AlMn TES. B: ACTPol MoCu TES. C: POLARBEAR-2 AlMn TES, $\sim$10 times smaller in area than the other TES films. D: AdvACT AlMn test TES, 16.5 $\mu$m wide. E: AdvACT AlMn test TES, 37.5 $\mu$m wide. F: location of AlMn film. Device parameters and geometries are listed in Table 1. (b): Resistance versus temperature plot for a TES at 10 $\mu$A excitation current when exposed to various values of magnetic field. $T_\mathrm{c}$ is determined for each applied field value and plotted as a function of magnetic field in Fig. 2.}
  \vspace{-1.5em}
\end{figure}

We tested TESes from ACTPol chips \citep{Grace2014}, AdvACT 150 GHz (HF) chips, AdvACT 30 GHz (LF) chips \citep{Duff2016}, TES test chips with AlMn films of varying geometries, and POLARBEAR-2 TES test chips \citep{Suzuki2016} with varying leg lengths for magnetic sensitivity. Two TESes of each type were measured. All AlMn films had concentrations of 2000 ppm per atomic $\%$ \citep{Li2016}. The POLARBEAR-2 TESes had thicknesses of 60 nm, and thus were higher $R_N$ and $T_\mathrm{c}$ devices than the 400 nm thick ACT TESes \citep{Schmidt2011,Li2016}. We chose excitation currents for the four-lead measurements to balance noise reduction in the measurements with minimizing power dissipation through the TES bolometers (Table 1). A current of 10 $\mu$A was selected for the lower $R_N$ devices, while a current of 100 nA was selected for the higher $R_N$ devices. Any heating of the devices due to the selected excitation current was minimal and not observed to significantly affect $T_\mathrm{c}$. For each device and at each applied magnetic field value, we took $T_\mathrm{c}$ to be the temperature value at 50$\%$ $R_N$, where $R_N$ is the resistance value measured 2 mK above the last superconducting data point in the resistance versus temperature curve at zero applied magnetic field. A plot of $T_\mathrm{c}$ versus applied magnetic field for the tested bolometers is shown in Fig. 2 along with 
parabolic fits to the points. Parameters from the parabolic fits to one of each type of device are listed in Table 1. The error bars on $T_\mathrm{c}$ are chosen to be 1.3 mK, the standard deviation of a Gaussian fit to the differences in recorded $T_\mathrm{c}$ between 18 otherwise identical data points taken over the course of two separate cooldowns for the AdvACT LF chips.

\begin{figure}
\centering
\includegraphics[width=1.0\linewidth,keepaspectratio]{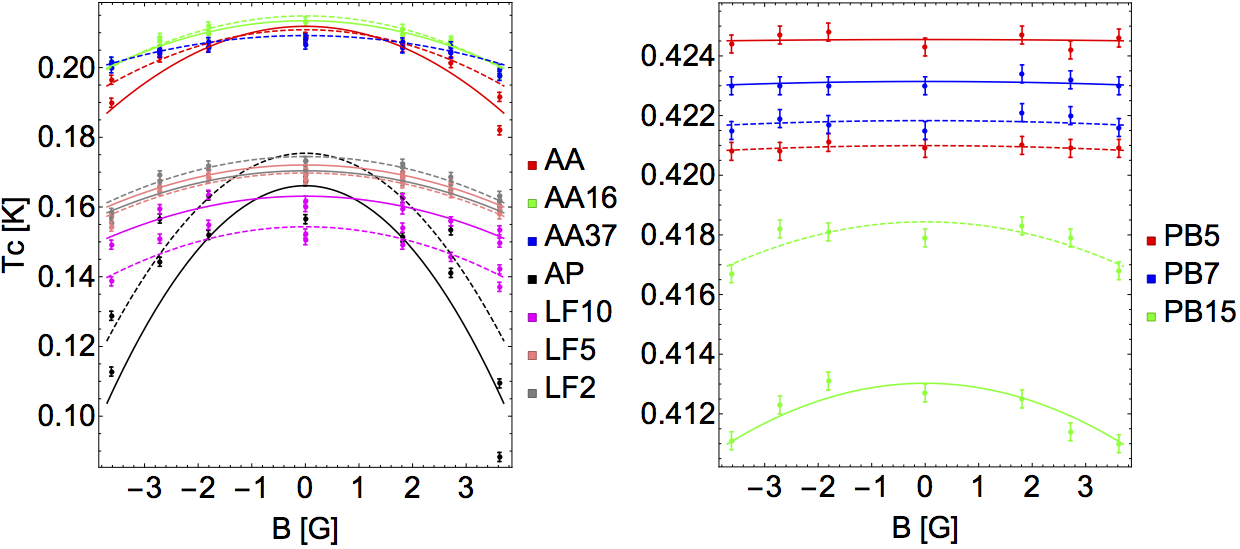}    
\centering
\caption{$T_\mathrm{c}$ versus $B$ for tested TESes (AdvACT (AA), ACTPol (AP), POLARBEAR-2 (PB), AdvACT LF (LF)) and parabolic fits to the data points. Dashed and solid lines indicate measurements of different devices of the same type. Coefficients from parabolic fits are listed in Table 1.}
\label{fig:figure2}
\end{figure}


\begin{table}
\begin{center}
\begin{tabular}{ |c|c|c|c|c|c|c| } 
 \hline
 TES &  \shortstack{\\ Leg $w$ x $l$ \\ $[\mu$m$]$ } & \shortstack{\\ AlMn Area, \\ Thickness} & R$_N[\Omega]$ & $T_{C} [K]$ & $w[\frac{K}{G^2}]$ & $\frac{dI_0}{dB} [\frac{\mu A}{G}]$ \\ 
 \hline
PB5           & 4,~10 x 500   & 610 $\mu $m$^2$, 60 nm  & 0.84    & 0.425 & 3.0e-6 & 4e-5\\ 
PB7           & 4,~10 x 700   & 610 $\mu $m$^2$, 60 nm  & 0.86    & 0.423 & 8.5e-6 & 5e-5\\ 
PB15          & 4,~10 x 1500  & 610 $\mu $m$^2$, 60 nm  & 0.86    & 0.416 & 1.1e-4 & 5e-4\\ 
AdvACT& 15 x 61  & 6200 $\mu $m$^2$, 400 nm & 0.007 & 0.211 & 1.2e-3 & 1e-1\\ 
AA16          & 20 x 61    & 3300 $\mu $m$^2$, 400 nm & 0.004 & 0.213 & 1.0e-3 & 1e-1\\ 
AA37          & 20 x 61    & 11250 $\mu $m$^2$, 400 nm & 0.007 & 0.208 & 0.6e-3 & 1e-1\\ 
AA LF         & 10 x 1000  & 6200 $\mu $m$^2$, 400 nm & 0.007 & 0.154 & 1.1e-3 & 3e-2\\
AA LF         & 10 x 500   & 6200 $\mu $m$^2$, 400 nm & 0.006 & 0.170 & 0.9e-3 & 3e-2\\
AA LF         & 10 x 220   & 6200 $\mu $m$^2$, 250 nm & 0.006 & 0.170 & 0.9e-3 & 5e-2\\ 
AP MoCu& 20 x 61          & 9000 $\mu $m$^2$, 250 nm & 0.007 & 0.166 & 4.6e-3 & 5e-1\\
\hline 
\end{tabular}
\caption{AdvACT (AA), ACTPol (AP), and POLARBEAR-2 (PB) TESes, leg lengths, AlMn (or MoCu for ACTPol) areas, and excitation currents, with parabolic fits in the form of $y = T_{C,B=0}-wx^2$ to $T_\mathrm{c}$ versus $B$ data for one of each type of tested TES. AA16 indicates an AdvACT test TES with a width of 16.5 $\mu$m, and AA37 indicates a width of 37.5 $\mu$m. PB TESes have one leg which is 4 $\mu$m wide and one which is 10 $\mu$m wide. All concentrations of AlMn were 2000 ppm per atomic $\%$. The high $R_N$ devices were tested with a 100 nA excitation current, while the low $R_N$ devices were tested using a 10 $\mu$A excitation current. Errors on parameters are taken to be 1.3 mK due to scatter in otherwise identical data points during separate cooldowns, and 20$\%$ of sensitivity fit $mK/Gauss^2$ due to fitting error. Estimates of $dI_0/dB$ should be regarded as comparative figures only and are based on an approximation of the sensitivity at B = 0.05 Gauss as described in Sect.~5.}
\vspace{-1.5em}
\end{center}
\label{table:table1}
\end{table}

\section{\label{sec:section2}Weak-Link-Like Behavior in AlMn TESes}
\vspace{1em}

A theoretical model of the physics governing the superconducting phase transition of TES bolometers has yet to be constructed. Experiments have shown that the critical current of square thin-film TESes depends upon the TES geometry and temperature which can be described in terms of longitudinal proximity effects in the weak-link model of TES films \citep{Sadleir2010}. The critical current of these TESes has been observed to show Fraunhofer-like oscillations in applied magnetic fields, similar to those observed in Josephson junctions \citep{Sadleir2010,SadleirThesis}. A Ginzburg-Landau model can be used to explain measurements of $I_\mathrm{c}(T)$ for TESes considered to be SN'S proximity induced weak-links, measured in bath temperatures near $T_\mathrm{c}$ \citep{Sadleir2010,SadleirThesis,Smith2013}. These measurements have previously been made for MoAu and MoCu bilayers, among others \citep{BennetandUllom}. In this work, we present observations of weak-link-like behavior in AlMn TESes. 

Using the same experimental field setup described in Sect.~1, magnetic fields are applied perpendicular to the plane of the TES films. The three TES devices tested were most similar to the AdvACT HF TESes (B. in Fig. 1). The TESes are read out using the same TDM readout system used in AdvACT with NIST SQUIDs similar to those in \citep{Henderson2016}. At each value of applied magnetic field or each value of temperature, we perform voltage ramps to get a reading of the critical current $I_\mathrm{c}$ at which the TES transitions from superconducting to normal. Applied magnetic field was ramped from 0 to positive applied field and from 0 to negative applied field as defined by the normal direction of the TES film. Varying the method of ramping magnetic flux was not observed to have a significant effect on $I_\mathrm{c}(B)$.

Plots of $I_\mathrm{c}$ versus $B$ are shown for three devices in Fig. 3. These data were acquired for the TESes at bath temperatures near $T_\mathrm{c}$ where the Ginzburg-Landau model would apply for the AlMn films. A plot of $I_\mathrm{c}$ versus T for the three devices is shown in Fig. 3 along with fits to the data where the Ginzburg-Landau model applies. The fits take the form $I_\mathrm{c}(T)=a\sqrt{T/T_\mathrm{c}-1}e^{-b\sqrt{T/T_\mathrm{c}-1}}$, where $a$ is proportional to the width of the device film and $b$ is proportional to the length \citep{Sadleir2010}. We observe a trend in $a$ consistent with the theory, with $a = 0.50 \pm 0.01 \times 10^{6} \ \mu$A, $b = 73 \pm 121$ for the 16.5 $\mu$m wide by 200 $\mu$m long AlMn device (``TES 2") and $a = 1.00 \pm 0.05 \times 10^{6} \ \mu$A, $b = 69 \pm 28$ for the 25 $\mu$m wide by 200 $\mu$m long devices (``TES 1" and ``TES 3").

\begin{figure}
  \centering
  \subfloat[]{\includegraphics[width=0.49\linewidth,keepaspectratio]{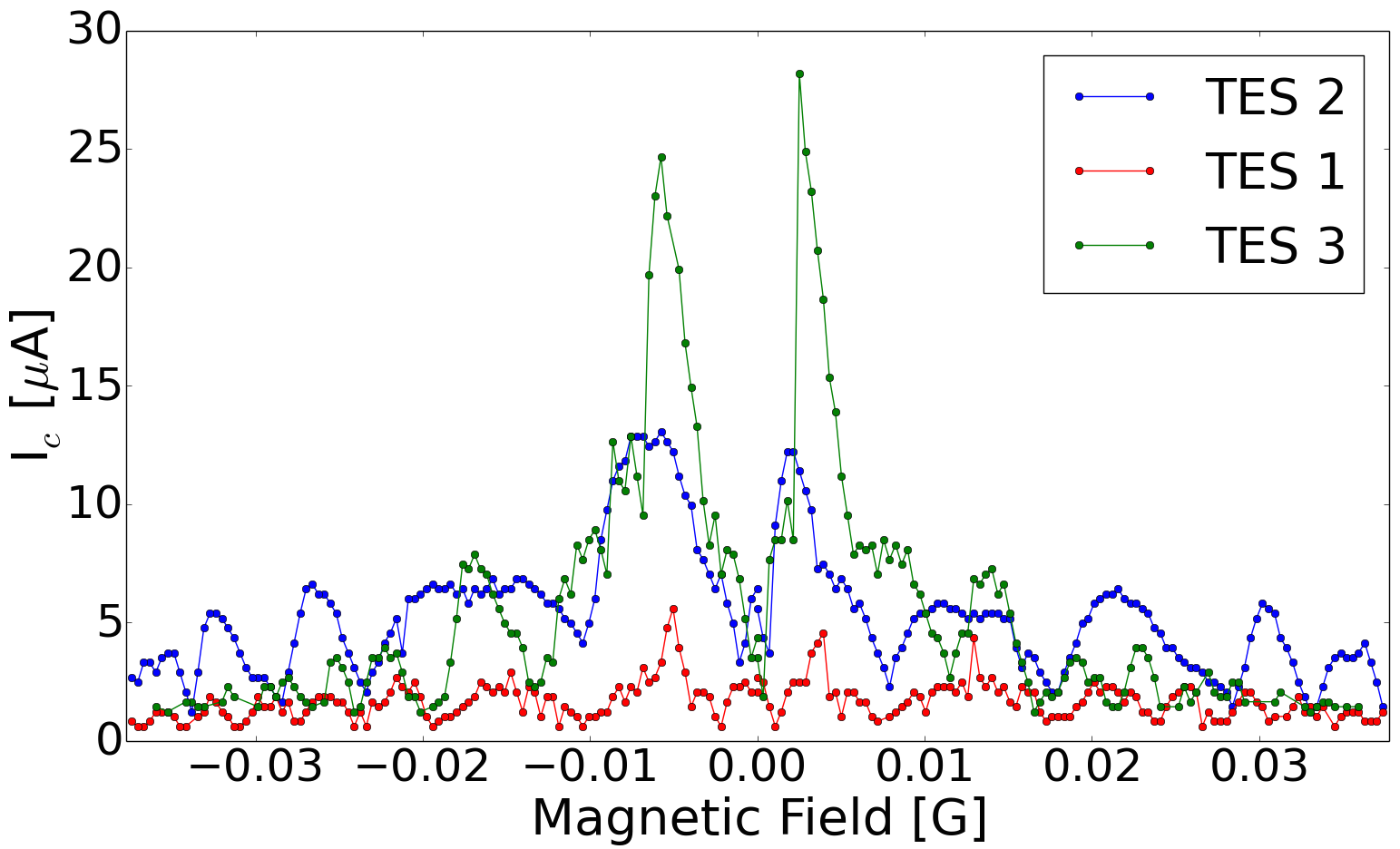}\label{fig:f1}}
  \hfill
  \subfloat[]{\includegraphics[width=0.50\textwidth]{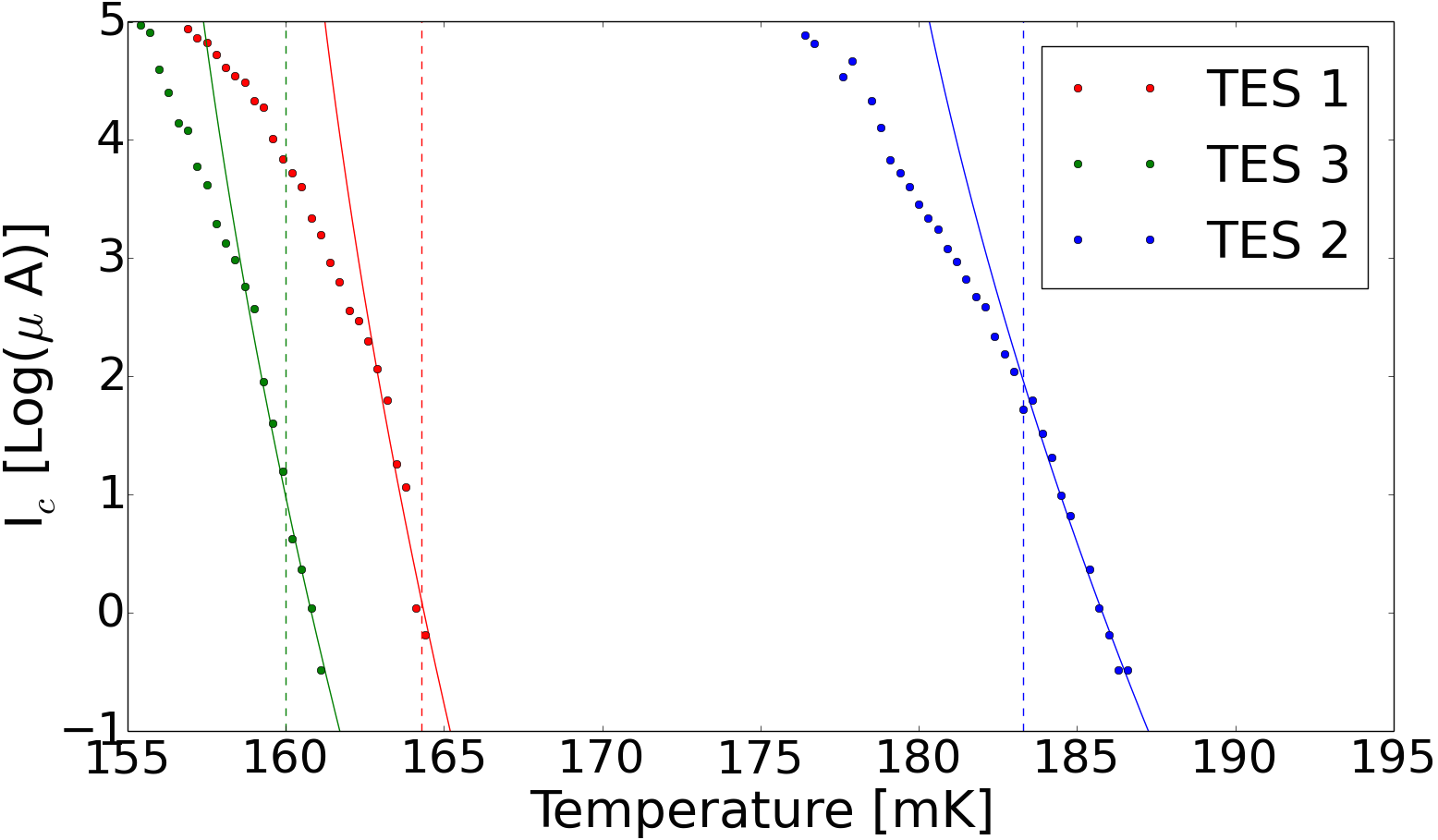}\label{fig:f2}}
  \caption{(a): $I_\mathrm{c}$ versus $B$ for three AlMn TES films, similar to B in Fig. 1. We observe Fraunhofer-like oscillations in the data. The period of the oscillations is irregular, as is their decay; however, the period is repeatable both between cooldowns and as we adjust the temperature along the trajectories shown to the right. The lack of a central peak requires further study. (b): $I_\mathrm{c}$ versus T for the three AlMn TES films, along with fits to the high temperature data using the Ginzburg-Landau model (equation 1). Temperature is held fixed for the measurements of $I_\mathrm{c}$ versus $B$ in (a) for each device, and is indicated here by the dashed vertical lines in (b).}
  \vspace{-1.5em}
\end{figure}

The behavior observed in these devices generally agrees with the weak-link model. We observe Fraunhofer-like oscillations in all three tested devices; however, the observed oscillations are not consistent in period or decay, and the absence of the central peak in the oscillations requires further study. The measured high temperature data are fit by the Ginzburg-Landau model and are consistent with expectations for the TES geometries studied here.

\section{\label{sec:section4}Magnetic Sensitivity of $\mu$MUX and TDM SQUIDs}
\vspace{1em}

To measure the magnetic sensitivity of the TDM SQUIDs described in Sect.~1, magnetic fields were applied perpendicular to the planes of the SQUIDs. The SQUIDs were mounted in the same MUX board used to read out AdvACT single pixels on the DR's coldest stage. V-$\phi$ curves were acquired for applied various field values using the MCE readout electronics. The shift in the V-$\phi$ curves due to the presence of positive and negative applied fields was measured for 411 readout channels (Fig. 4).  By calculating $d\phi_0$/$\phi_0$ per Gauss for each channel and considering the average distribution of these values, we determine the upper bound on measured TDM SQUID sensitivities to be 1.2 $\phi_0$/Gauss.     

To estimate the magnetic sensitivity of $\mu$MUX rf-SQUIDs, magnetic fields were applied perpendicular to the planes of 33 rf-SQUIDs on a single NIST $\mu$MUX 14a chip and time-ordered data were taken on each rf-SQUID using a ROACH readout system, returning an average phase response of the $\mu$MUX channel in radians as a function of applied magnetic field. Data were taken for two different orientations of the chip within the magnetic shield (Fig. 4). A gradient in response to the magnetic field was seen across the $\mu$MUX chip in the first orientation, with a minimum in sensitivity at the central rf-SQUIDs and maxima at the ends of the chip (Fig. 4). This slope is thought to be due to the sensitivity of the gradiometric winding of the SQUID coils to gradients in magnetic field as a function of position inside the DR, since the same response was not observed in the second orientation of the chip within the shield. We place an upper limit on magnetic sensitivities of 0.3 $\phi_0$/Gauss for the $\mu$MUX rf-SQUIDs.

\begin{figure}
  \centering
  \subfloat[]{\includegraphics[width=0.52\linewidth,keepaspectratio]{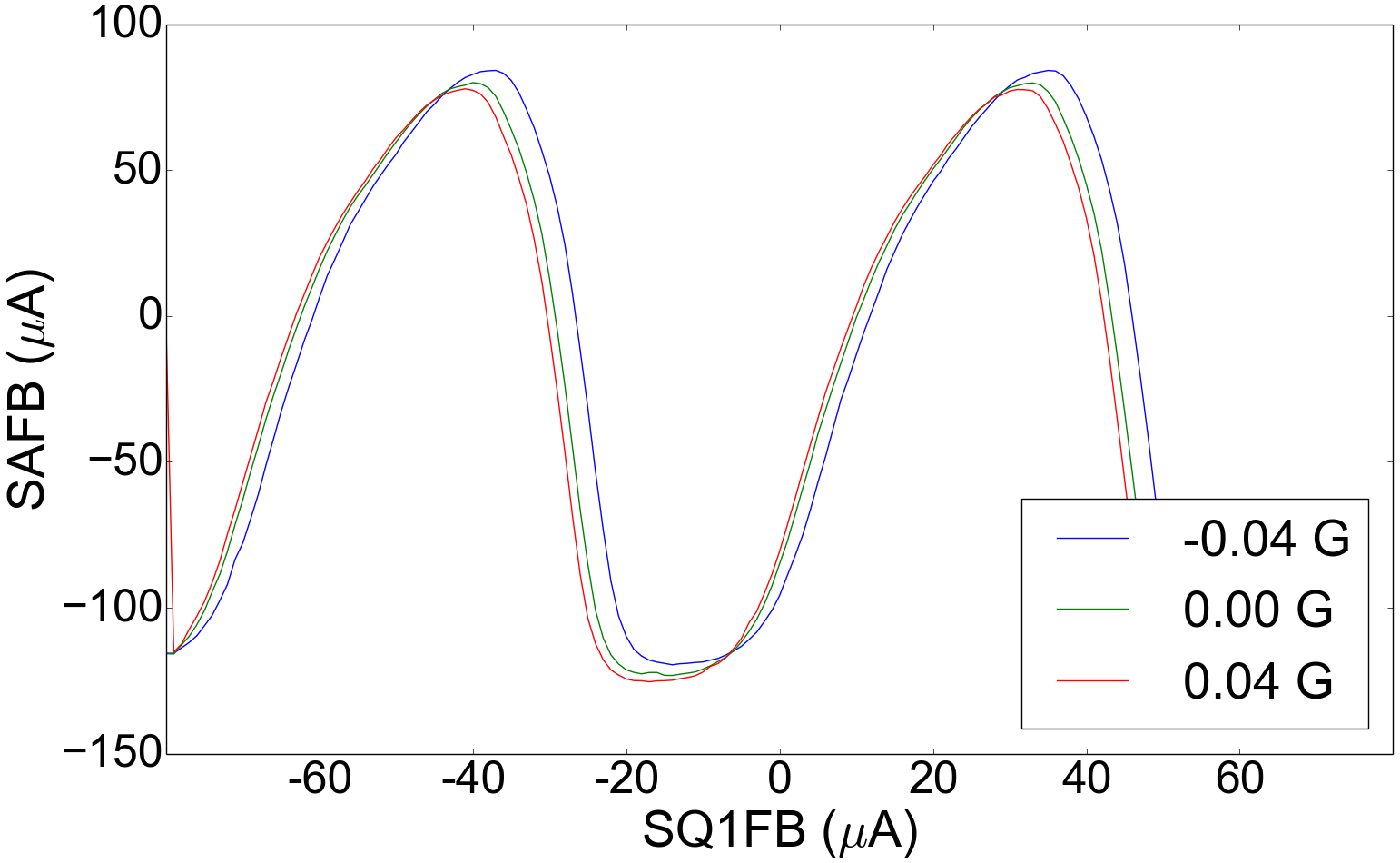}\label{fig:f1}}
  \hfill
  \subfloat[]{\includegraphics[width=0.48\textwidth]{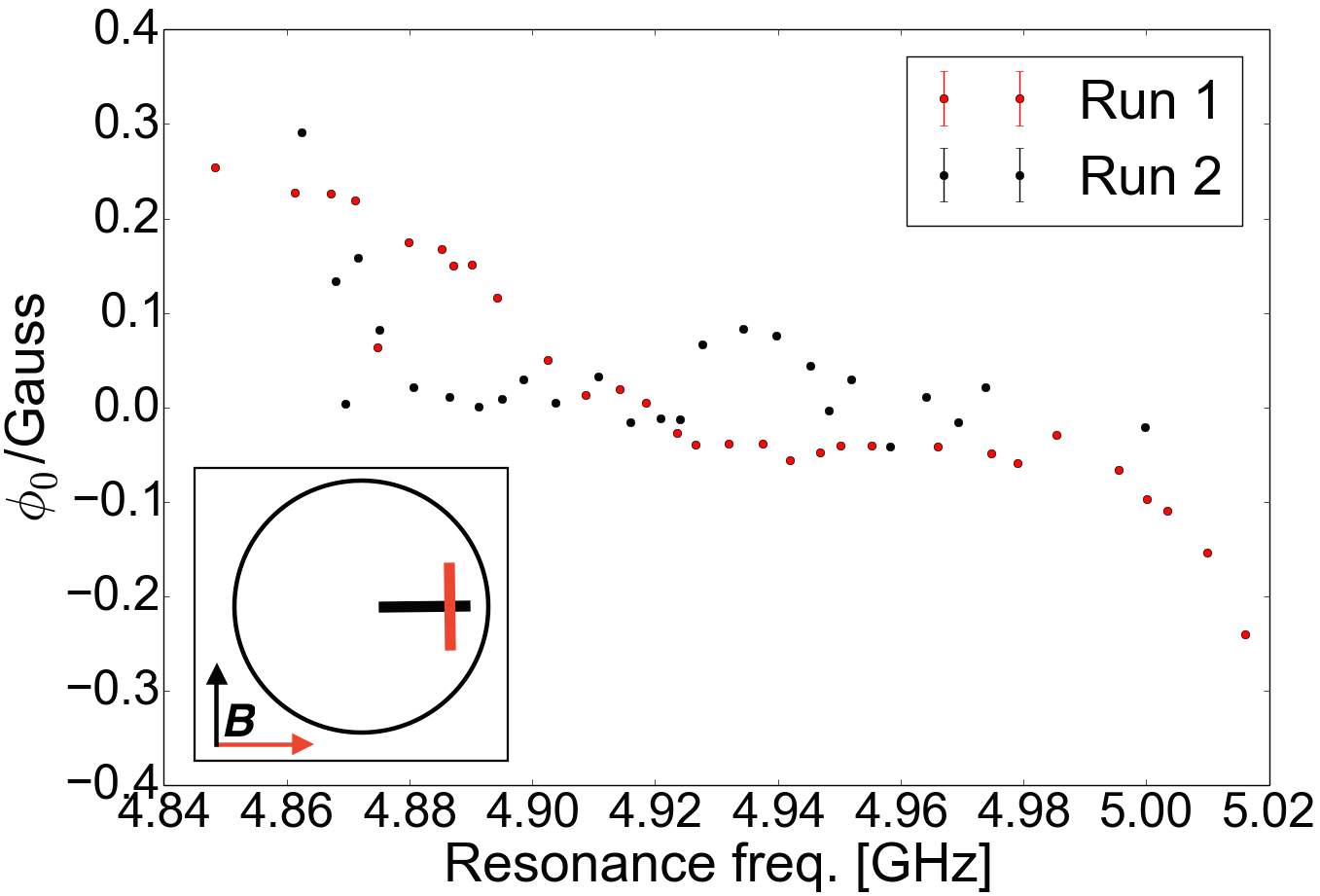}\label{fig:f2}}
  \caption{(a): An example of shifts in the V-$\phi$ curve of a single TDM SQUID under the influence of applied magnetic field. (b): $\phi_0$/Gauss for resonances on the $\mu$MUX chip display a gradient in response across the chip for the Run 1 orientation (red) but not for the Run 2 orientation (black). The top view schematic diagram shows the position of the $\mu$MUX chip within the magnetic shield for the two runs along with the applied magnetic field directions outside the shield. The chip resonances run from low to high frequency from top to bottom in the red orientation, and left to right in the black orientation in this view. The number of channels read out and the accompanying central frequencies varied slightly between the two runs. An upper limit on the magnetic sensitivities of these rf-SQUIDs is taken to be 0.3 $\phi_0$/Gauss.}
  \vspace{-1.5em}
\end{figure}

\section{\label{sec:section5}Shielding Considerations for Future CMB Experiments}
\vspace{1em}

Magnetic shielding designs for upcoming CMB experiments should be driven by device sensitivities such as those presented in this work in order to minimize cost and extent of mechanical design. Using the measurements obtained for our tested TESes and SQUIDs, we can convert detector and readout magnetic sensitivities into estimates of the change in detector bias current per applied magnetic field by using $\frac{\delta I_0}{\delta B} \approx \frac{G(T_\mathrm{c}) (-2wB)}{V_0}$, where $w$ is our parameter fit listed in Table 1, $B$ is a magnetic field value offset from zero (taken to be 0.05G, or $\sim$1/10 Earth's magnetic field) and G is the thermal conductance of the TES \citep{NiemackThesis}. Using appropriate values for the types of TESes tested, we obtain sensitivity estimates in detector bias current and list them in Table 1 \citep{Patty2016,GraceThesis,SuzukiThesis,Koopman2017}. Because these sensitivities are estimated at an arbitrary value of magnetic field, and the true relationship between $dB$ and $dI_0$ is more complex than fully represented in this estimate, these numbers should be treated as a comparative guide to relative sensitivities.  

A similar calculation can be done to convert the TDM SQUID sensitivity estimate into a predicted detector current response as a function of magnetic fields inside the shielding, using conversion factors particular to our readout setup \citep{NiemackThesis}. For our upper limit sensitivity, 1.2 $\phi_0$/Gauss, we estimate $\delta I_{0 \ \mathrm{eff}}/\delta B \approx 100 $\\$ \ \mu \mathrm{A} /\mathrm{Gauss}$, three orders of magnitude larger than the estimates for our TESes. For the $\mu$MUX rf-SQUIDS, with an upper limit sensitivity of 0.3 $\phi_0$/Gauss due to the gradiometric response of the rf-SQUIDs, $\delta I_{0 \ \mathrm{eff}}/\delta B \approx 4 \ \mu \mathrm{A} /\mathrm{Gauss}$, a factor of 25 times less sensitive than the TDM SQUIDs. 

In combination with knowledge and experience drawn from current experiments, this information will serve to motivate magnetic shielding designs for future CMB efforts. We have simulated magnetic shields with ANSYS Maxwell to estimate the magnetic shielding factors of existing shielding geometries as well as possible configurations for future experiments. We estimate an ACTPol style Amumetal 4K 32-cm-diameter double (single) layer cylindrical magnetic shield to have a shielding factor for on-axis fields of $\sim$500 (100) at the location and orientation of the AdvACT detector arrays and TDM SQUIDs. This factor appears to increase to $>$ 7500 for DC fields perpendicular to the cylinder's axis, which was the most sensitive orientation for the previous ACTPol TDM SQUIDs. The TDM SQUIDs used for ACTPol did not show any evidence of external magnetic field pickup as the cryostat was rotated through Earth's field.  Preliminary studies with AdvACT do not provide evidence for significant pickup, but more detailed AdvACT analysis is needed.  If we scale the less sensitive $\mu$MUX SQUID pickup levels from either the ACTPol or the AdvACT SQUID shielding factors, they suggest $\mu$MUX shielding factor targets between 20 and 300. For comparison, a shielding factor of $\sim$50 was achieved in SCUBA-2 by enclosing the experiment's SQUIDs in a niobium box within high-permeability shields on the inside of the vacuum vessel \citep{SCUBA2}. 

\section{\label{sec:lastsection}Conclusion}
\vspace{1em}

We have made measurements of the magnetic sensitivity of AlMn and MoCu TESes, varying in geometry, leg length and doping, TDM SQUIDs, and $\mu$MUX rf-SQUIDs. The MoCu ACTPol TESes are the most sensitive to magnetic fields, followed by the AdvACT AlMn TESes, with the POLARBEAR-2 AlMn TESes being the least sensitive. The primary source of the differences between the sensitivities of the AlMn TESes is not yet clear, though we note that the POLARBEAR-2 and AdvACT TESes do have significantly different areas, critical temperatures, doping, and thicknesses. An observation of weak-link-like behavior in AlMn TESes at low critical currents was made. Further study could help inform how this behavior impacts detector parameters. We used estimates for AlMn TES, TDM SQUID, and $\mu$MUX SQUID magnetic sensitivities from device measurements along with simulations to motivate realistic shielding factors that would sufficiently suppress field excursions in upcoming experiments. These results will inform the design of magnetic shielding for future CMB experiment receivers such as those for CCAT-prime, Simons Observatory, and CMB-S4 and thereby help enable precision measurements of the CMB sky.

\begin{acknowledgements}
The authors thank Christine Pappas for useful discussions of weak-link-like behavior in AlMn TESes, Zeqi Gu for assistance in measuring magnetic shielding values, and Suzanne Staggs, Edward Wollack, and Kevin Crowley for their helpful comments and feedback which have improved this work. The authors also thank the Atacama Cosmology Telescope, Simons Array, and Simons Observatory collaborations for their contributions, including the development of the detectors tested in this paper. This work was supported by NSF Grant AST-1454881. EMV was supported by the NSF GRFP under Grant No. DGE-1650441. 

\end{acknowledgements}

\vspace{1 em}

\bibliography{apj-jour,LTD17_Vavagiakis_RefResponse}{}
\bibliographystyle{ltd16}

\end{document}